# Memristor-based mono-stable oscillator

A.T. Bahgat and K.N. Salama

In this letter, a reactance-less mono-stable oscillator is introduced for the first time using memristors. By replacing bulky inductors and capacitors with memristors, the novel mono-stable oscillator can be an area-efficient solution for on-chip fully integrated systems. The proposed circuit is described, mathematically analysed and verified by circuit simulations.

*Introduction:* In light of HP lab's 2008 announcement [1], huge interest in memristors has revived and it has been investigated in many circuits and systems [2-5]

In this letter, a novel reactance-less mono-stable oscillator is proposed using memristors. The resistance storage property of the memristor replaces the energy storage elements needed for oscillation. The increase/decrease in the memristor resistance $R_m$ according to the polarity of the applied voltage is similar to charging and discharging of capacitors [3]. Memristor-based reactance-less mono-stable oscillator offers a solution in low frequency applications that require large capacitance area or off-chip reactive components.

*The proposed circuit:* The proposed circuit operation depends on the memristor resistance oscillation that is tracked by the memristor voltage $V_m$. The circuit consists of a voltage divider, a feedback system and two output terminals as shown in Fig. 1.

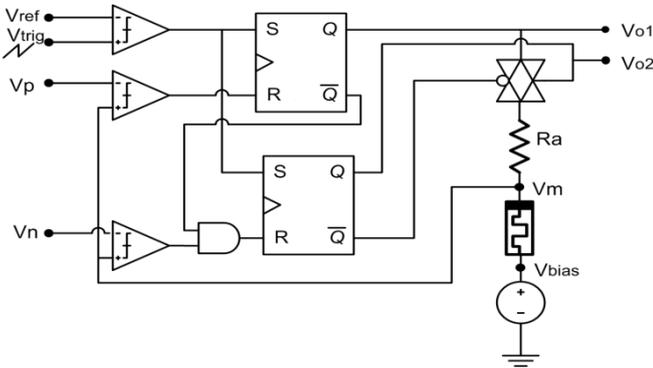

**Fig. 1** The proposed memristor-based mono-stable oscillator

The voltage divider network is formed between the memristor, a resistor $R_a$ and a small transmission gate resistance $R_{trans}$. The memristor is connected in a way that its resistance $R_m$ increases when $V_m$ is higher than $V_{bias}$ and vice versa. Since the transmission gate has very low resistance $R_{trans} \ll R_a$ and its resistance changes during the oscillation, $R_{trans}$ is ignored to simplify mathematical solutions.

$$V_m(t) = (V_o(t) - V_{bias}) \frac{R_m}{R_m + R_a + R_{trans}} + V_{bias} \qquad (1)$$

The feedback system consists of two comparators and an AND gate controlling two flip flops. Depending on Vm, the first flip flop controls the voltage connected to the voltage divider network and the second flip flop controls turning the transmission gate on and off. The two outputs of the flip flop are the two output terminals of the circuit.

*Circuit tracing:* Before any external trigger, the oscillator is in the stable state. The transmission gate is turned off and $R_m$ is preserved since no voltage drop across the memristor exists. After the trigger pulse, the behaviour of the circuit changes from 'stable' passing to other operation points and goes back to the 'stable' point as shown in Fig. 2.

**stable → a:** The oscillator enters the unstable operation points. The transmission gate is turned on as high output voltage $V_{oh}$ is applied on both outputs of the two flip flops $V_{o1}$ and $V_{o2}$. Then, $V_m$ switches its value to $V'_p$ according to (2) and the operating point jumps to (a).

$$(V_m - V_{bias}) = (V'_p - V_{bias}) = (V_n - V_{bias})\left(\frac{V_{oh} - V_{bias}}{V_{ol} - V_{bias}}\right) \qquad (2)$$

**a → b:** At point (a), the voltage difference across the voltage divider $(V_{oh} - V_{bias})$ is positive. Hence, the memristor resistance increases and accordingly, $V_m$ increases as well until it reaches $V_p$ and the operating point reaches (b).

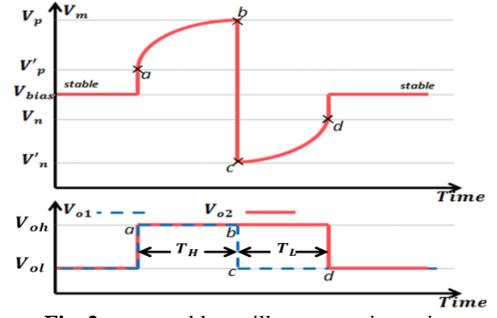

**Fig. 2** mono-stable oscillator operation points

**b → c:** At point (b), the value of $V_m$ just passes $V_p$. So, $V_{o1}$ switches to low output voltage, $V_{ol}$, and $V_m$ changes to $V'_n$ following (3).

$$(V'_n - V_{bias}) = (V_p - V_{bias})\left(\frac{V_{ol} - V_{bias}}{V_{oh} - V_{bias}}\right) \qquad (3)$$

**c → d:** At point (c), the voltage difference $(V_{ol} - V_{bias})$ is negative, so the memristor resistance decreases. $V_m$ increases until it reaches $V_n$ and the operating point reaches (d).

**d → stable :** At point (d), the value of $V_m$ just passes $V_n$ so $V_{o2}$ switches to low output voltage $V_{ol}$ turning off the transmission gate. This makes the voltage divider network floating and keeps the memristor resistance $R_m$ preserved. This ends of the oscillation cycle of the memristor resistance and the oscillator goes back to the stable state.

With each trigger pulse and resistance oscillation cycle, each output pulse has a certain expected pulse width except the first cycle because all cycles start and end with the memristor resistance except the first cycle which starts with the initial memristor resistance $R_o$.

*Oscillation condition:* In the voltage divider network, for each $V_m$, there is only one equivalent memristor resistance $R_m$. The memristor resistance at $V_m = V_p$ and $V_m = V_n$ is given by

$$R_{mp} = (R_a)\left(\frac{V_p - V_{bias}}{V_{oh} - V_p}\right) \quad , \quad R_{mn} = (R_a)\left(\frac{V_n - V_{bias}}{V_{ol} - V_n}\right) \qquad (4)$$

Based on circuit tracing, $V_{bias}$, $V_p$ and $V_n$ must be selected such that

$$V_p - V_{bias} > (V_n - V_{bias})\left(\frac{V_{oh} - V_{bias}}{V_{ol} - V_{bias}}\right) \qquad (5)$$

Since $R_{on} < R_{mn} < R_{mp} < R_{off}$, then from (4), the resistance $R_a$ is restricted by (6)

$$R_{on}\left(\frac{V_{ol} - V_n}{V_n - V_{bias}}\right) < R_a < R_{off}\left(\frac{V_{oh} - V_p}{V_p - V_{bias}}\right) \qquad (6)$$

Similar to other mono-stable oscillators, there must be minimum time between two consecutive trigger pulses. This minimum time in the proposed oscillator is the time required for the memristor to finish a complete resistance oscillation which is the second output pulse. In addition, the trigger pulse width must be less than the first output pulse width. The result of violating those two conditions is shown in Fig. 3.

*Output pulse width:* the proposed circuit has two output voltages, $V_{o1}$ and $V_{o2}$, with different output pulse widths, $T_{o1}$ and $T_{o2}$ in order. The output pulse width of $V_{o1}$ and $V_{o2}$ are determined by

$$T_{o1} = T_H = Time\,(R_{mn} \rightarrow R_{mp}) \qquad (7)$$

$$T_{o2} = T_H + T_L = T(R_{mn} \rightarrow R_{mp}) + T(R_{mp} \rightarrow R_{mn}) \qquad (8)$$

For the circuit mathematical analysis, a developed mathematical model for HP memristor is used to analyse the circuit mathematically [6,7]. The memristor resistance as a function of time is given by

$$R_m^2(t) = R_o^2 \pm 2K \int_0^t \Delta V_m(T)\,dT \qquad (9)$$

where the constant $K$ is defined as $K = \mu_v\,R_{on}(R_{off} - R_{on})/d^2$ and $\Delta V_m(T)$ is the voltage difference across the memristor. By solving the integration for the time required for the memristor resistance to increase from $R_{mn}$ to $R_{mp}$ and substituting into (1). Then $T_H$ is given by

$$T_H = \frac{R_{mp}^2 - R_{mn}^2 + 2(R_a)(R_{mp} - R_{mn})}{2K(V_{oh} - V_{bias})} \qquad (10)$$

Similarly, the time required for the memristor resistance to decrease from $R_{mp}$ to $R_{mn}$, $T_l$, is given by

$$T_L = \frac{R_{mn}^2 - R_{mp}^2 + 2(R_a)(R_{mn} - R_{mp})}{2\,k(V_{ol} - V_{bias})} \quad (11)$$

From the equations (4), (10) and (11), the ratio between the widths of the two output pulses, $V_{o1}$ and $V_{o2}$, is determined by

$$\frac{T_{01}}{T_{02}} = \frac{V_{bias} - V_{ol}}{V_{oh} - V_{ol}} \quad (12)$$

*Circuit simulation*: As the access to the experimental realisation of the memristor is very limited, researchers resort to SPICE and behavioural models of the HP memristor [8], or emulate the memristor model using active circuitry [4]. The proposed circuit was simulated and the memristor parameters, $R_{on}$, $R_{off}$, $d$ and $\mu_v$, were chosen to be $100\Omega$, $38k\Omega$, $10nm$ and $10^{-14}\,cm^{-1}s^{-1}V^{-1}$ respectively [3,7]. Circuit parameters, $R_a$, $V_p$, $V_n$, $V_{bias}$, $V_{ol}$ and $V_{oh}$, were choosed to be $8K\Omega$, $0.8V$, $0.3V$, $0.5V$, $0V$ and $1V$ respectively. Fig. 3 shows the transient simulation results showing excellent matching with the mathematical analysis. The memristor resistance oscillates between Rmn and Rmp, which defines the location of the operating points. By substituting the memristor and the circuit prammeters into (4), (10) and (11) : $R_{mn} = 5.3k\Omega$, $R_{mp} = 12k\Omega$, $T_{01} = 0.58$ sec and $T_{02} = 1.176$ sec.

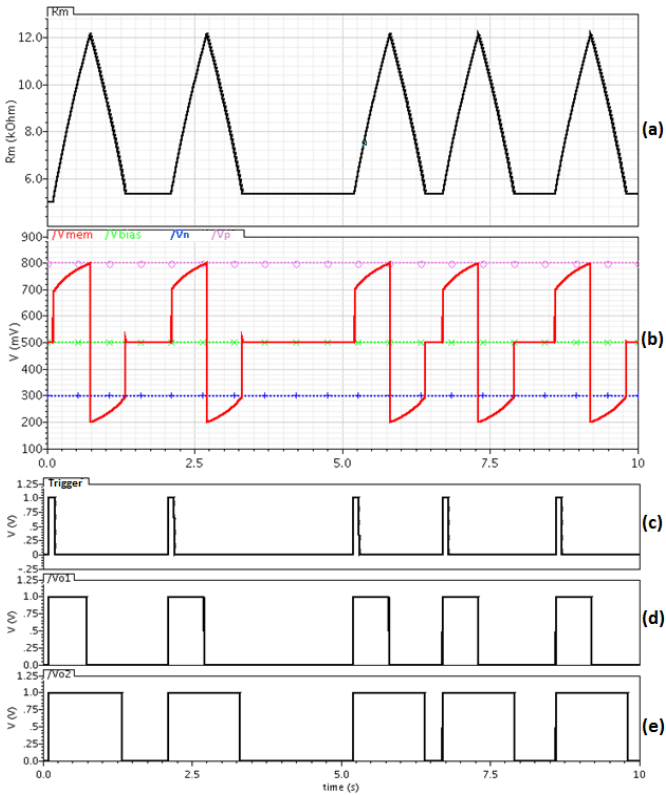

**Fig. 3** the behavior of the mono-stable oscillator. (a) shows memristor resistance, $R_m$, oscillation (b) shows $V_m$ changing between $V_p$, $V_n$ and $V_{bias}$ (c) shows trigger pulse (d) shows $V_{o1}$ pulses (e) shows $V_{o2}$ pulses

Fig.4 shows the effect of $R_a$ on the width of the output pulses. Simulations were run choosing $V_p = 0.9V$, $V_n = 0.4V$ and $V_{bias} = 0.5V = 0.5\,V_{cc}$ as values for our circuit parameters. Both simulations and calculations match with an error of approximately 2%. This error is expected because the resistance of the transmission gate is ignored to simplify the mathematical expression.

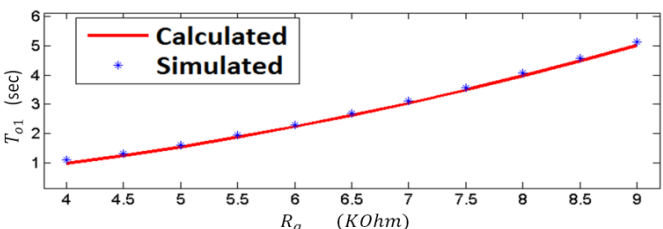

**Fig.4** simulated and calculated first output pulse width vs. Ra

*Conclusion:* This paper proposes a novel reactance-less mono-stable oscillator with a detailed mathematical analysis that is verified by circuit simulation. The design is suitable for low frequency applications. Further, the proposed circuit is general and can be extended to higher frequencies.

A.T. Bahgat and K.N. Salama (King Abdullah University of Science and Technology (KAUST), Thuwal 23955-6900, Saudi Arabia)

E-mail: ahmed.bahgat@kaust.edu.sa